\documentclass[preprint,12pt]{elsarticle}

\usepackage{amsmath,amssymb}
\usepackage{lineno}
\modulolinenumbers[5]

\journal{Journal of Magnetism and Magnetic Materials}

\begin{document}

\begin{frontmatter}

\title{Magnetic Constitution of Topologically Trivial Thermoelectric {PbTe:Cr} }

\author[1]{Katarzyna Gas\corref{mycorrespondingauthor}}
\cortext[mycorrespondingauthor]{Corresponding author: Katarzyna Gas}
\ead{kgas@ifpan.edu.pl}

  \address[1]{Institute of Physics, Polish Academy of Sciences, Aleja Lotnikow 32/46, PL-02668 Warsaw, Poland.}
\address[2]{International Research Centre MagTop, Institute of Physics, Polish Academy of Sciences, Aleja Lotnikow 32/46, PL-02668 Warsaw, Poland}

\author[1]{Aleksandra Krolicka}
\author[1,2]{Krzysztof Dybko}
\author[1]{Piotr Nowicki}
\author[1]{Zeinab Khosravizadeh}
\author[1,2]{Tomasz Story}

\author[1]{Maciej Sawicki}

\begin{abstract}
In this paper we report on detailed temperature and magnetic field dependence of magnetization of IV-VI semiconductor PbTe doped with mixed valence transition metal Cr$^{2+/3+}$.
The material is studied solely by a SQUID magnetometer in order to quantitatively determine the contribution of single substitutional  Cr$^{2+/3+}$ as well as of various Cr-Te magnetic  nanocrystals, including their identification.
The applied experimental procedure reveals the presence of about  $10^{19}$~cm$^{-3}$  paramagnetic  Cr ions, of which 2/3 are the Cr$^{3+}$ ions formed via self-ionization of Cr$^{2+}$ resonant donors.
 These are known to improve the thermoelectric figure of merit parameter zT of this  semiconductor.
The magnetic finding agrees with previous  Hall effect studies thus providing a new experimental support for the proposed electronic structure model of PbTe:Cr system with resonant  Cr$^{2+/3+}$ state located (at low temperatures) about 100 meV above the bottom of the conduction band.
Below room temperature a ferromagnetic-like signal points to the presence of Cr-rich nanocrystalline precipitates.
Two most likely candidates, namely: Cr$_2$Te$_3$ and  Cr$_5$Te$_8$ are identified upon dedicated temperature cycling of the sample at the remnant state.
As an ensemble, the nanocrystals exhibit (blocked) superparamagnetic properties.
The  magnetic susceptibility of both n- and p-type PbTe in the temperature range $100 < T < 400$~K has been established.
These magnitudes are essential to properly account for the high temperature magnetic susceptibility of PbTe:Cr.
\end{abstract}

\end{frontmatter}

\section{Introduction}

Composed of very heavy elements PbTe is known for strong relativistic (spin-orbital) effects in the band structure resulting in narrow band gap (mid infrared range) and very good thermoelectric properties. In the complete series of PbTe-SnTe solid solutions PbTe is a topologically trivial terminal compound with SnTe possessing an inverted band ordering and constituting a topological crystalline insulator \cite{Dziawa:2012_NM,Tanaka:2012_NP,Xu:2012_NComm}. Topological transition can be induced in PbTe by applying hydrostatic pressure or by mixing with other semiconductor materials.

The performance of the ever-so-much important thermoelectric energy conversion is described by the material thermoelectric figure of merit (zT).
Since its improvement in nanostructured materials due to the impeding of the phonon heat flow has neared the amorphous materials fundamental limit, alternative ways of band gap engineering towards the enlargement of the density of states at the Fermi level \cite{Mahan:1996_NAS}, e.g. via the band convergence \cite{Pei:2011_N} or a co-doping to induce a resonant level in the electric transport relevant band \cite{Heremans:2008_S}, are considered.

In the world-wide search for new thermoelectric energy converting materials  magnetic semiconductors and semimetals constitute a distinguished class of materials that offer additional degrees of freedom in controlling electric and thermal transport effects.
These, in particular, concern the following effects: the exchange splitting of energy bands induced by the sp-d exchange coupling between charge carriers and local magnetic moments, the magnon transport contribution to thermal and thermoelectric effects, the introduction of  spin-dependent energy selective carrier scattering mechanisms, and shaping electron density of states in semiconductors by incorporation of transitions metal deep donor or acceptor centers (see e.g.~refs.~\cite{Heremans:2008_S,Ahmed:2017_JMCA,Mori:2017_Small,Acharya:2018_JMCC,Bhat:2019_MTP,Zheng:2019_SA,Tsujii:2019_SA,Wiendlocha:2018_PRB}).
Some of these new effects are expected in systems exhibiting long-range magnetic order whereas the other are also important in nonmagnetic materials doped with paramagnetic ions (the case studied in this work).

Doping PbTe with magnetic ions of Cr has several  thermoelectrically  important aspects.
Cr in PbTe is an  efficient n-type dopant with  special properties (expected very good  electrical homogeneity) related to its mixed valence 2+/3+ charge state.
The energy level of Cr$^{2+/3+}$ state is energetically located in very close proximity to Fermi level what results in  beneficiary (increased thermoelectric power) modification of electron density of states and  introduction of efficient energy-selective resonant carrier scattering mechanism (as proposed, e.g.~in ref.~\cite{Heremans:2008_S}).
In the latter respect, PbTe has acquired a status of the model example where zT doubled after introduction 1-2\% of Tl \cite{Heremans:2008_S}.
Thermoelectric properties of PbTe can be also controlled by incorporation of 3d transition metals, such as Mn, Fe or Cr \cite{Story:2002_B, Biplab:2011_APL,Skipetrov:2014_APL,Dybko:2016_APL,Krolicka:2017_PB-CM,Krolicka:2017_CrystRes}.
Chromium is particularly interesting since Cr substituting for Pb atoms form a mixed valence Cr$^{2+/3+}$ donor centers resonant with the PbTe conduction band \cite{Biplab:2011_APL,Mac:1995_APP,Akimov:1989_FTP,Vulchev:1986_pss,Nielsen:2012_PRB}.
In contrast Cr in SnTe is expected to form level resonant  with the valence  band \cite{Skipetrov:2009_PhysB}.
The Cr$^{3+}$ states are formed due to the self-ionization of Cr$^{2+}$ ions.
This fact was confirmed by the observation of the magnetically active Cr$^{3+}$ ions in electron paramagnetic resonance spectrum \cite{Story:1993_APP}.
The electrons generated alongside fill the conduction band up to about 100 meV, the approximate position of both the Cr$^{2+/3+}$ and the Fermi levels.
However, the solubility limit of Cr in IV-VI compounds is rather limited, 0.1 to 0.5\%, depending on the exact growth conditions, so a nanocrystalline material is typically obtained. It consists of a paramagnetic host dilute magnetic semiconductor (DMS): (Pb,Cr)Te, which is enriched by various Cr-rich nanometer-sized crystallites.
Therefore, the magnetic response of Cr-doped samples is complex, but the substantially different magnetic characteristics of these constituencies allow to separate them out and quantify.
On the other hand, the presence of nanoinclusions may profit in thermoelectric performance enhancement by a considerable reduction of thermal conductivity \cite{Biswas:2012_N}.

In this report we present results of a complete magnetic analysis of a typical PbTe:Cr sample grown by the Bridgman method \cite{Mac:1995_APP,Story:1992_APP}.
Our aim is to determine the concentration of electrically and magnetically active substitutional Cr ions serving as donor centers resonant with the conduction band in PbTe, what is important for thermoelectric properties.  Magnetometric methods constitute an experimental tool best suited for this task, although the analysis of experimental observations is obscured by a complicated  magnetic constitution of  (Pb,Cr)Te ingots grown by the Bridgman method, as shown in this work.

\section{Samples and experimental details}

\begin{figure}[b!]
	\begin{center}
		\includegraphics[width = 8.0cm]{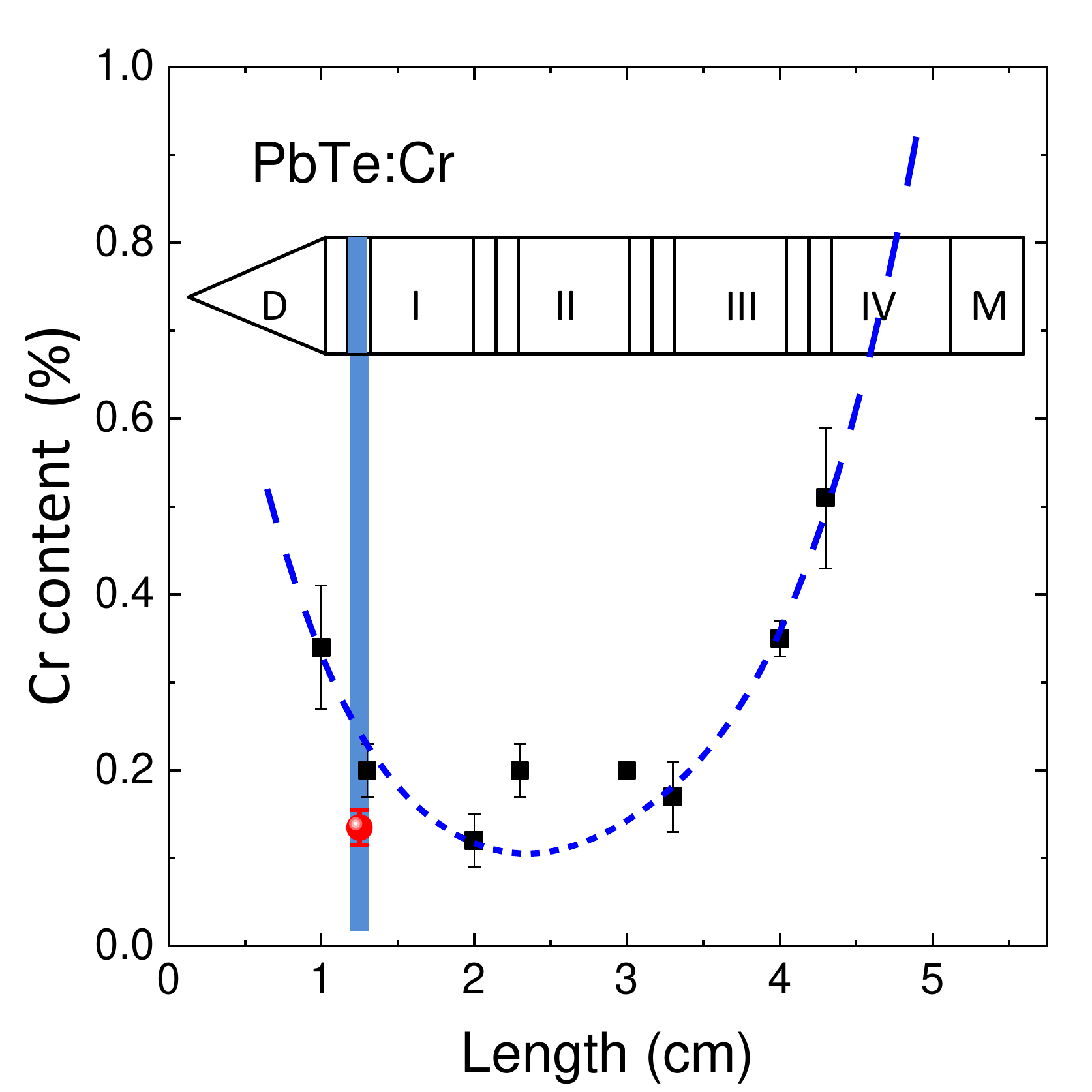}
		\caption{Distribution of Cr content along PbTe:Cr ingot grown by the Bridgman method. The solid squares are the results of the energy dispersive x-ray fluorescence analysis. The error bars show differences between two sides of 1~mm-thick discs wire-cut from the ingot depicted in the figure. The dashed line is a guide for the eyes. The origin of the sample investigated in this study is marked by the thick bar. The red bullet marks the total Cr contents established upon the magnetic studies reported below.}
		\label{Fig-sklady}
	\end{center}
\end{figure}
We investigate crystalline PbTe:Cr samples wire-cut from a large crystal grown by the modified Bridgman method    using pre-synthetized polycrystalline PbTe ingot and metallic Cr of a purity better than 5N.
For the material investigated here the Cr content $x$ varies non-monotonously along the ingot, as determined by x-ray fluorescent analysis. This is depicted in Fig.~\ref{Fig-sklady}.
Chromium exhibits a very low solubility in major semiconductors systems, so precipitates of other Cr-rich nanocrystals (cf.~eg.~ref.~\cite{Karczewski:2003_JSNM}) in PbTe are very likely too, and, in fact, were ubiquitously reported \cite{Mac:1995_APP,Nielsen:2012_PRB,Zvereva:2010_JPCS}.

It has been observed previously \cite{Mac:1995_APP,Story:1992_APP} that the number, chemical composition and size of the Cr-rich precipitates increase as one moves up towards the end of the ingot.
Therefore, to keep this contribution at minimum the  specimen detailed  here originates  from the beginning of the ingot.
The approximate location of the origin of the sample is marked by the thick bar in Fig.~\ref{Fig-sklady}.
We confirm that other neighboring specimens  exhibit semi-quantitatively identical results.
In the course of thorough characterization  of our PbTe:Cr samples we experimentally studied at room temperature the Seebeck effect, the Hall effect, and the electrical conductivity confirming expected electron concentration of $n \simeq 7 \times 10^{18}$~cm$^{-3}$ and a large thermoelectric power $\alpha = -(220-295)$~$\mu$V/K.
We estimate the room temperature magnitude of zT parameter of our PbTe:Cr to range between 0.2 and 0.25 [for a realistic value of the thermal conductivity 1.5~W/(m$\cdot$K)].
An $n$ close to $10^{19}$~cm$^{-3}$ indicates that the Fermi level is pinned to the resonant Cr$^{2+/3+}$ state formed by the self-ionization of substitutional Cr$^{2+} \longrightarrow$ Cr$^{3+} + e^-$, and that two kinds of Cr ions are expected to be seen \cite{Story:1993_APP,Kossut:1990_SST}.

Magnetic measurements are performed using an MPMS XL Superconducting Quantum Interference Device (SQUID) magnetometer equipped with a low field option at magnetic fields $H$ up to 70 kOe in the temperature $T$ range between 2 and 400~K.
Prior to the zero-field studies such as the thermoremnant magnetization a slow degaussing followed by a soft quench of the SQUID's superconducting magnet (the "magnet reset" option of the MPMS magnetometer) is performed.
Approximately $2 \times 2 \times 1$~mm$^3$ specimen is  glued by a minute amount of a nonmagnetic GE varnish to the sample holder made of about 20~cm long Si stripe.
To eliminate experimental artifacts we strictly follow the experimental code and data reduction detailed in ref.~\cite{Sawicki:2011_SST}.


\section{Results and discussion}

\begin{figure}[t!]
\begin{center}
\includegraphics[width = 14cm]{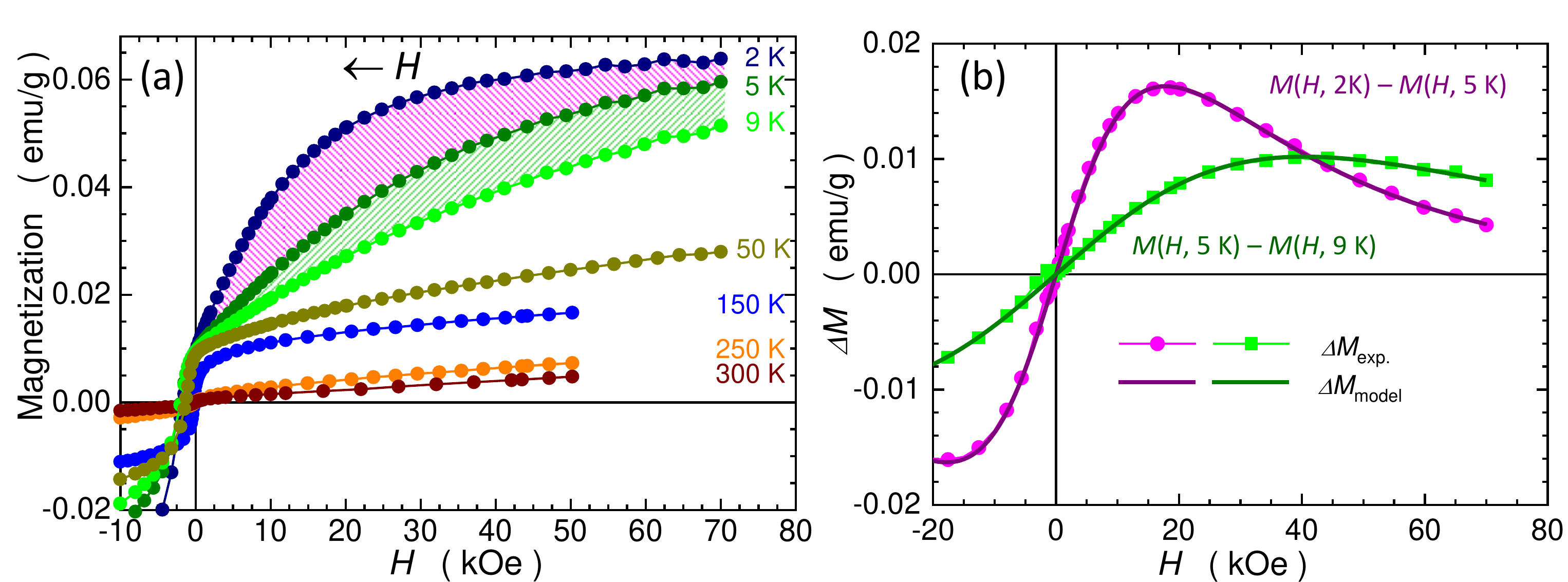} 
\caption{(Color online) (a) Magnetic isotherms $M(H)$ at selected temperatures after subtraction of the PbTe host lattice diamagnetism (bullets, the lines are guides to the eye). For clarity of the presentation only one way sweeps of the magnetic field $H$ are shown. The hatched areas indicate the differences of $M(H)$ analyzed in (b). (b) Magnitudes of experimental differences of $ M(H)$, $\Delta M_{\mathrm{exp}}(H)$  established between $T=2$ and 5~K (magenta bullets) and 5 and 9~K (green squares). The solid lines, $\Delta M_{\mathrm{model}}$, represent the sum of calculated differences of two Brillouin functions $g \mu_{\mathrm{B}} x_{\mathrm{Cr}} J \times \Delta B_{J}(T,H)$ at the same pairs of temperatures for $J=3/2$ and $J=2$  for $x_{\mathrm{Cr^{3+}}} =7.8(5) \times 10^{18}$~cm$^{-3}$ and $x_{\mathrm{Cr^{2+}}} = 3.5(6) \times 10^{18}$~cm$^{-3}$, respectively. }
\label{Fig-MH}
\end{center}
\end{figure}
The main result of the present report is shown in Fig.~\ref{Fig-MH}.
A representative suite of the $M(H)$ curves measured at temperatures between 2 and 300~K  is presented in Fig.~\ref{Fig-MH}~(a).
The $T$-dependent diamagnetic contribution of PbTe lattice  has been subtracted.
We detail on this issue in the further parts of this report.
Clearly two contributions can be identified: a ferromagnetic-like (FM) response developing below 250~K  and a slowly saturating paramagnetic (PM)  component, dominating at very low temperatures.
We can quantitatively evaluate the strength of this PM  part employing the concept elaborated previously for GaN:Fe \cite{Pacuski:2008_PRL} since the data presented in Fig.~\ref{Fig-MH}~(a) are qualitatively similar to $M(H)$ curves seen in other material system in which this evaluation method proved to work very well \cite{Navarro-QA:2010_PRB,Navarro-QA:2020_Materials,Gas:2021_JMMM_NWs}.

To evaluate the concentration of the PM ions, $x_{\mathrm{PM}}$, we investigate the differences of the $M(H)$ isotherms, $\Delta M_{\mathrm{exp}}(H)$ between 2 and 5~K and 5 and 9~K, as indicated by hatched areas in Fig.~\ref{Fig-MH}~(a) and presented by symbols in Fig.~\ref{Fig-MH}~(b).
The applicability of this procedure stems from  a rather high magnitude of the spin-ordering temperature of the FM-like part, so its contribution to  the magnetization can be regarded as $T$-independent at very low temperatures.
However, a small complication appears here since two kinds of substitutional Cr ions coexist in the specimen.
All substitutional Cr ions assume "initially" Cr$^{2+}$  ($3d^4$) configuration, but those which self-ionize convert to Cr$^{3+}$ ($3d^3$)  one.
In the octahedral coordination of PbTe the ground level of the Cr$^{3+}$  ions at zero magnetic field is an orbital singlet ($L=0$, $S=3/2$).
The magnetic moments of these ions result solely from the spin, so their $H$- and $T$-dependent properties are described by the Brillouin function $B_J(T,H)$, $J=S=3/2$.
The ground state of the non-self-ionized Cr$^{2+}$  ions ($L=2$, $S=2$) is an orbital doublet which undergoes the Jahn-Teller distortion, so the resulting ground state is again an orbital singlet whose magnetic response can also be described by $B_J(T,H)$ and the Land\'e factor $g=2.0$, but with $J=S=2$  \cite{Ono:1954_PR}.
This leaves us with only two adjustable parameters, namely the concentrations of both Cr species, $x_{\mathrm{Cr^{3+}}}$ and $x_{\mathrm{Cr^{2+}}}$.
We obtain these two magnitudes by fitting $\Delta M_{\mathrm{exp}}(H)$ by the sum of two differences of  the $B_J(T,H)$ functions calculated separately for each $J=3/2$ and $J=2$ at the temperatures specified above, namely by the sum of two contributions of the form of $g \mu_{\mathrm{B}} x_{\mathrm{Cr}} J \times \Delta B_J(T,H)$, where $\mu_{\mathrm{B}}$ is the Bohr magneton.
The solid lines in Fig.~\ref{Fig-MH}~(b) are obtained for   $x_{\mathrm{Cr^{3+}}} = 7.8(5) \times 10^{18}$~cm$^{-3}$ and $x_{\mathrm{Cr^{2+}}} = 3.5(6) \times 10^{18}$~cm$^{-3}$. The sum of these concentrations amounts to $1.13(8) \times 10^{19}$~cm$^{-3}$, or $x_{\mathrm{PM}}/N_0 \simeq 0.08$\%, where $N_0 = 4/a^3$  is the concentration of cation  sites in the rock salt crystal lattice with the lattice parameter $a=6.46$~\AA.

Having evaluated the concentration of the substitutional Cr ions we calculate their contribution to $M$ at any given $T$ and $H$, $M_{\mathrm{PM}}(T,H)$, and subtract these values from the experimental data to establish magnitudes of the magnetization of the FM-like part of the material, $M_{\mathrm{FM}}(T,H)$.


\begin{figure}[t!]
\begin{center}
\includegraphics[width = 8.0cm]{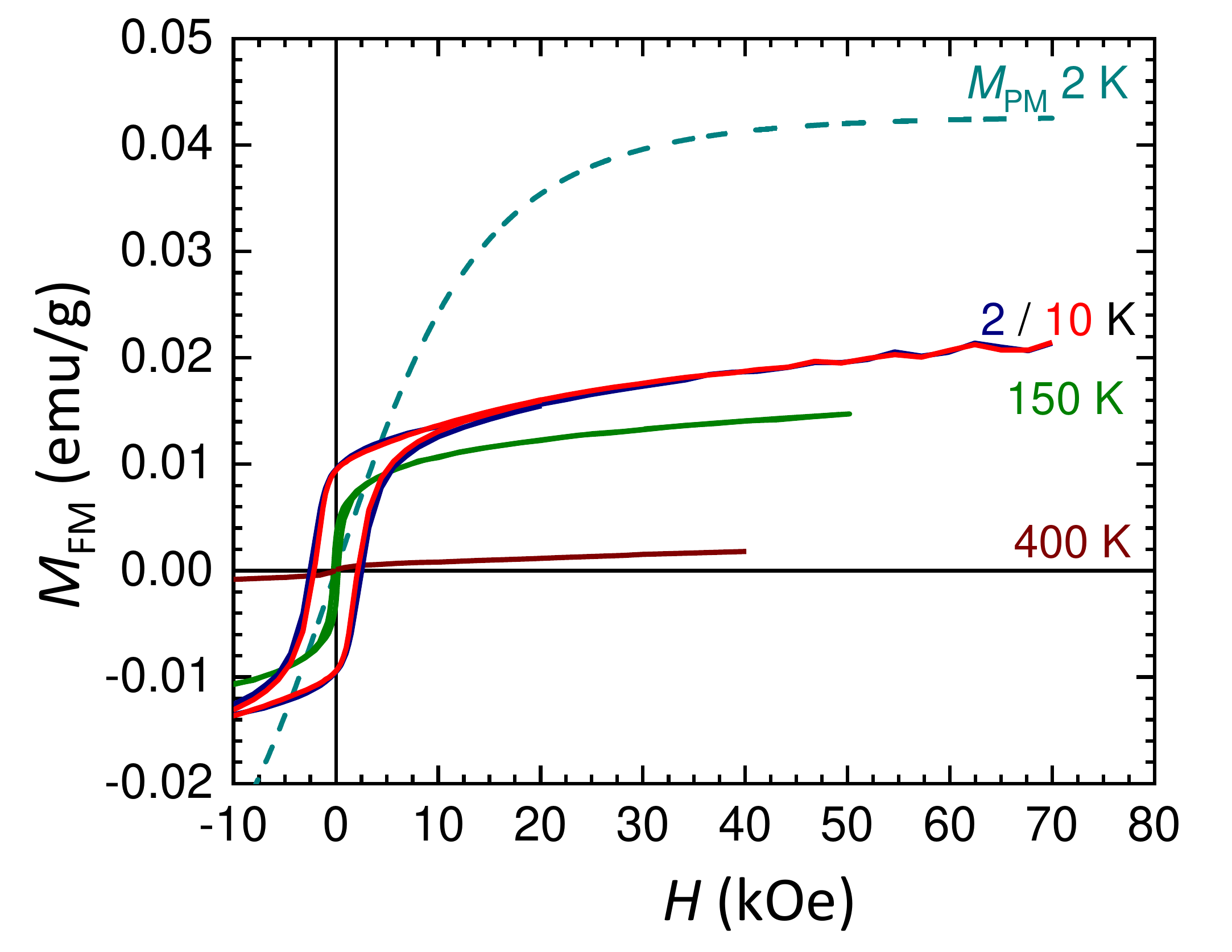}
\caption{(Color online) The ferromagnetic-like contribution to magnetization, $M_{\mathrm{FM}}$, plotted as a function of the external field $H$ - solid lines.  The dashed cyan line represents the paramagnetic contribution to magnetization, $M_{\mathrm{PM}}$, calculated at 2 K for the parameters established by the method illustrated in Fig.~\ref{Fig-MH}~(b) }
\label{Fig-FM}
\end{center}
\end{figure}
The bare  FM-like contribution is denoted in Fig.~\ref{Fig-FM} by solid lines.
The exemplified there  hystereses of $M_{\mathrm{FM}}(H)$ are rounded, i.e. more "S-like" than rectangular and the values of the remnant magnetization are very close to 50\% of the corresponding saturations.
This about 50\% ratio is the first magnetic indication of the granular form of $M_{\mathrm{FM}}$.
Indeed, such a magnitude of remnant magnetization is expected for an ensemble of randomly oriented uniaxial magnetic particles  \cite{Stoner:1948_PTRSL}.

The dashed line in Fig.~\ref{Fig-FM} denotes the magnitude of $M_{\mathrm{PM}}$ calculated at $T=2$~K for $x_{\mathrm{PM}} = 1.13 \times 10^{19}$~cm$^{-3}$, as established above.
The magnitude of the saturation of $M_{\mathrm{PM}}$ allows us to estimate  the  relative strength of both $M_{\mathrm{PM}}$ and $M_{\mathrm{FM}}$ contributions.
It reads about 2:1 in favor for the PM one.
This finding allows to make a coarse estimation of the total Cr contents in the sample to about 0.12\%.


\begin{figure}[t!]
\begin{center}
\includegraphics[width = 8.0cm]{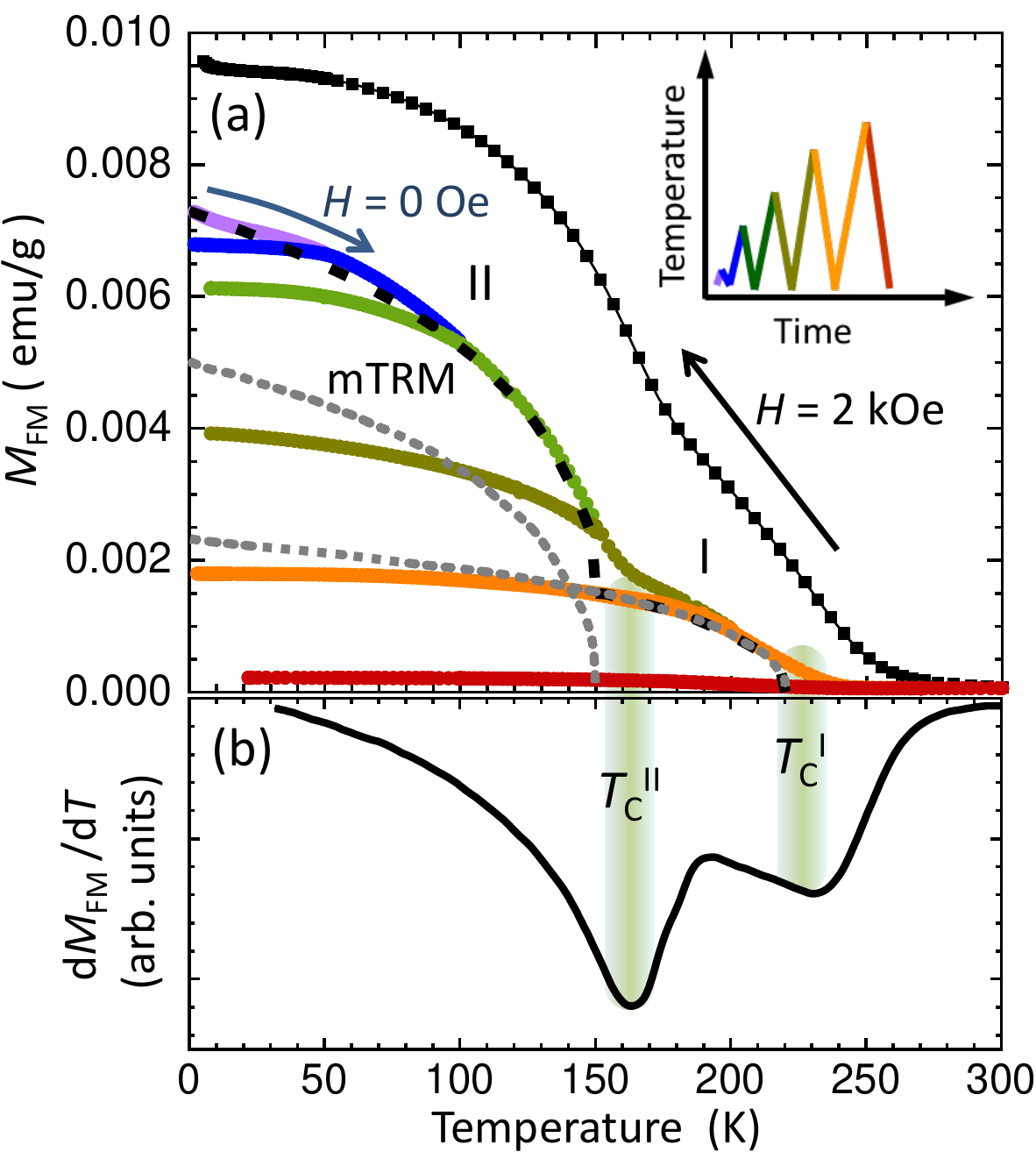}
\caption{(Color on line) (a) Temperature dependence of magnetization specific to the ferromagnetic-like part of the sample, $M_{\mathrm{FM}}$, recorded at magnetic field $H= 2$~kOe (black squares), and the  thermoremnant magnetization measured in an oscillating manner, mTRM (lines). The inset shows the thermal history of the sample during the mTRM measurement. The two dotted gray lines mark the two major contributions to mTRM, approximated by the Langevin functions $L(T)$. $L(T)$ describes the temperature dependence of the spontaneous magnetization obtained in the classical limit ($J = \infty$) of the molecular field theory. The thick dashed line is the sum of these two $L(T)$. (b) First derivative of $M_{\mathrm{FM}}(T)$ from panel (a). The most likely magnitudes of Curie temperatures for the two contributions responsible for the two-step shape of $M_{\mathrm{FM}}(T)$ and mTRM are marked by thick shaded bars.}
\label{Fig-MT}
\end{center}
\end{figure}
More information on the magnetic constitution of $M_{\mathrm{FM}}$ can be inferred from  $T$-dependent studies in weak magnetic fields, summarized in Fig.~\ref{Fig-MT}~(a).
A two-step shape of  $M_{\mathrm{FM}}(T)$ acquired at $H = 2$~kOe, black squares, indicates that the FM-like signal originates from two components, differing with magnitudes of their spin-coupling temperature, $T_{\mathrm{C}}$.
We label these components in Fig.~\ref{Fig-MT}~(a)  as I and II and their coupling temperatures as $T_{\mathrm{C}}^{\mathrm{I}}$  and  $T_{\mathrm{C}}^{\mathrm{II}}$, respectively.
Accordingly, $T_{\mathrm{C}}^{\mathrm{I}} >  T_{\mathrm{C}}^{\mathrm{II}}$.
Both these magnitudes can be assessed by the differentiation of the $M_{\mathrm{FM}}(T)$ curve with respect to $T$.
The numerically obtained derivative $dM_{\mathrm{FM}}/dT$ is presented in Fig.~\ref{Fig-MT}~(b).
It points to $T_{\mathrm{C}}^{\mathrm{I}} \simeq  230$~K and $T_{\mathrm{C}}^{\mathrm{II}} \simeq  160$~K.
The same two components can be recognized on the $T$-dependence of the remnant magnetization, however, this vital measurement is performed in a modified way (mTRM), as purposed by Mamiya \emph{et al}.~\cite{Mamiya:2007_JMMM}.

The main modification within mTRM is that instead of continuous warming at $H=0$ of the beforehand pre-cooled sample in a saturating field, the warming is performed in an oscillating manner, as sketched in the inset to Fig.~\ref{Fig-MT}~(a).
It reads that after warming up the sample to a certain temperature, it is cooled back to the base temperature and warmed again to a higher one.
The process of warming and cooling cycles continues until the total signal from the sample vanishes.
Such a measurement suite allows to distinguish between the spontaneous magnetization in the equilibrium state in the zero field conditions of a bulk FM from a decaying with temperature one  \cite{Mamiya:2007_JMMM}.
The latter is due to the dynamical slow down by energy barrier(s) - a process specific to superparamagnetic (SP) materials below the so called blocking temperature.
Therefore, in blocked SP materials a step-like  dependence of mTRM is expected, 
whereas in bulk FM mTRM trails the one universal $T$-dependence  of its spontaneous magnetization.
In the latter case mTRM acquires a concave curvature, specific to the shape of the relevant Brillouin function $B_J(T)$.
Interestingly, the simultaneous presence of these two different forms of mTRM allow also to quantify the strength or the volumetric fraction  of SP and FM phases when both coexist in the specimen \cite{Sawicki:2010_NP,Gas:2021_JMMM_NWs}.

We start the analysis of the results of mTRM from a notion that the last part of this exercise, the cooling down from $T=300$~K$> T_{\mathrm{C}}^{\mathrm{I}}$. There is practically no signal recorded in this leg of the mTRM (its residual non-zero values are due to the presence of a sub-Oe residual magnetic field in the bore of the superconducting magnet of the magnetometer).
Such a contrast between the results obtained during the cooling at $H=2$~kOe and $H \simeq 0$ means that the ferromagnetic coupling below $T_{\mathrm{C}}^{\mathrm{I}}$ is caused by either spatially separated objects - precipitates exhibiting strong magnetic moments (macrospins), or by a homogeneous ferromagnet in a multi-domain state, or both.
In fact the FM ordering has been ubiquitously reported in various semiconductors which were incrusted with magnetic ions.
This field has become particularly active after the inspiring paper predicting even the room temperature FM mediated in such systems by itinerant carriers  \cite{Dietl:2000_S} via the Zener mechanism \cite{Zener:1951_PRa}.
However, the underlying conditions for the formation of a homogeneous (a "wall-to-wall") \emph{high temperature} coupling in DMS require a simultaneous presence of a high \emph{hole} concentration (i.e. exceeding $10^{20}$~cm$^{-3}$) and a higher than 5\% substitutional concentration of randomly distributed magnetic ions.
The FM coupling is certainly possible for much smaller concentrations of the magnetic elements, via the Ruderman-Kittel-Kasuya-Yosida interaction \cite{Ruderman:1954_a,Yosida:1957_a}, but then the coupling temperatures fall into the single Kelvins realm, or even below \cite{Cohrane:1974_PRB}.

None of the conditions specified above is even remotely fulfilled in the studied sample, but, it should be underlined here, PbTe-SnTe mixed crystals enriched with Mn up to 3\% were the very first DMS in which the presence of carrier driven homogeneous ferromagnetic phase has been established \cite{Story:1986_PRL}.
This exclusively low-$T$ feature has been observed only in p-type samples and the highest Curie temperatures (but "only" about 33 K) were observed for $p \simeq 3\times 10^{21}$~cm$^{-3}$ and Mn concentration approaching 15\% \cite{Lazarczyk:1997_JMMM_169,Lazarczyk:1997_JMMM_176}.
Similarly, somewhat low coupling temperatures are typically observed in as grown (Ga,Mn)As - the flag ship of the dilute ferromagnetic semiconductors family.
All possible technological advances, resulting in Mn concentration in excess of 10\% and hole concentration exceeding $10^{21}$~cm$^{-3}$ after the low temperature annealing treatment \cite{Edmonds:2004_PRL}, have not really promoted $T_{\mathrm{C}}$ to break the 200~K ceiling  \cite{Jungwirth:2005_PRB(72)}, arriving typically at about 175~K for the majority of the best optimized materials \cite{Sawicki:2006_JMMM}.
The material studied here is an \emph{n-type} DMS.
Here, in magnetically homogeneous n-type samples, the itinerant-electrons-driven FM coupling has been hinted only in sub-Kelvin temperatures \cite{Andrearczyk:2001_ICPS}.

Most certainly, the carrier-driven mechanism is not the only form of a possible coupling.
The short range superexchange mechanism can operate in insulating media.
But in such a case, a percolating paths connecting the majority of magnetic ions have to form.
For the nearest neighbors case this is limited to some 20\%.
This prohibiting  condition in DMS can be substantially weakened when the interaction with more distant ions is also FM, as it is in (Ga,Mn)N \cite{Simserides:2014_EPJ-WoC}, where a uniformly orchestrated coupling is  possible for more dilute cases \cite{Stefanowicz:2013_PRB}.
However, it takes place only at the low end of the cryogenic temperatures \cite{Sawicki:2012_PRB,Kunert:2012_APL}.
On the other hand, and importantly, it has been known that even tiny variations from the optimal growth condition can result in modification of the resulting  magnetic constitution of the DMS material, leading to both structural and magnetic phase separation.
In (Ga,Mn)N DMS only a tiny 10~$^{\mathrm{o}}$C difference in the growth temperature of about 600~$^{\mathrm{o}}$C results in modifications of the incorporation of Mn as large as 40\% \cite{Gas:2018_JALCOM}.
Bigger deviations lead to the formation of alien magnetic phases, breaking down the premise of the uniform magnetic systems \cite{Navarro-QA:2010_PRB,Bonanni:2010_CSR,Dietl:2010_NM}.
These are the most common cases where the FM-like response is observed at room temperature.
Most characteristically  these phase separated systems exhibit (i) a (very) low specific magnetization, ranging down to a tiny fraction of $\mu_B$ per magnetic atom, when the strength of the magnetic response is related to the whole volume of the sample and (ii) the lack of a monotonic dependence of all of their experimental magnetic properties, like the coupling temperature, on the concentration of the magnetic element.

Yet another possibility for the FM coupling in DMS provides the double exchange mechanism.
It does look promising in PbTe:Cr compound, since, indeed, depending on the Cr content, a mixed valence Cr$^{2+}$/Cr$^{3+}$ ensemble has been found in our sample with the occupation ratio of about 2:1 in favor of Cr$^{3+}$.
However, even this optimistic scenario requires a certain minimal concentration of magnetic ions to become effective, particularly at the elevated temperatures. For example, more than 2\% of Mn is needed to form the FM phase by the double exchange in  (Ga,Mn)N, but a peak value of $T_{\mathrm{C}}$ of only 100~K is expected for about 10\% of Mn \cite{Sato:2010_RMP}.

Lastly, all of these coupling mechanisms have  to be sufficiently strong to overcome the competing short-range antiferromagnetic superexchange occurring between the majority of the magnetic ions species in DMS \cite{Furdyna:1987_JAP}.
We can therefore summarize  these considerations by stating that a uniform weak ferromagnetism at elevated temperatures is very unlikely to form here and so we assign the lack of the spontaneous magnetization and the double step feature of the low field $M(T)$ curves between 150 and 250~K to the presence of two different species of Cr-rich precipitates, that is exclusively to the existence of a magnetically granular component existing in the otherwise paramagnetic PbTe:Cr matrix.

The conclusion formulated above about the granular form of  $M_{\mathrm{FM}}$ is further substantiated experimentally by the whole trace of mTRM data presented in Fig.~\ref{Fig-MT}~(a).
The experimental lines  (collected according to the temperature variations shown in the inset) form a unique pattern of a ladder-like structure expected to be seen in blocked SP materials [cf. Fig.~2~(b) in ref.~\cite{Mamiya:2007_JMMM}].
Such a response is caused by  the dynamical slowdown of the macrospins and corroborates the SP-like nature of  $M_{\mathrm{FM}}(T)$, i.e.  the FM-like features presented in Figs.~\ref{Fig-MH}~--~\ref{Fig-MT}.

However, there is a noticeable departure from the ideal picture presented in ref.~\cite{Mamiya:2007_JMMM}.
Instead of flat horizontal steps in the ladder-like mTRM as seen there, we record a great deal of a concave  curvature characteristic for a \emph{bulk} FM probed in a $T$-range encompassing its $T_{\mathrm{C}}$.
This rounding  indicates that the  Curie temperatures of the precipitates present in our sample are smaller than the whole $T$-range of mTRM performed here.
In turn it means that the exemplary horizontal steps of the mTRM ladder have to roll down at their high-$T$ sides to reflect the detrimental role of $T$ on the magnitudes of the macrospins' moments, when the sample temperature approaches $T_{\mathrm{C}}$ of a particular subset of the Cr-rich precipitates.
Therefore, this is only a much wider relative $T$-range of our studies that differentiates the results obtained here from the exemplary case discussed in ref.~\cite{Mamiya:2007_JMMM}, since $T_{\mathrm{C}}$ of the Fe$_3$N nanocrystals investigated there exceeded by far the $T$-range of the investigations: $T < 50$~K~$ << T_{\mathrm{C}}^{\mathrm{Fe}_3\mathrm{N}} \simeq 650$~K  \cite{Leineweber:1999_JALCOM,Navarro-QA:2020_Materials}.

The whole structure of mTRM allows also an independent estimation of the statistically most relevant magnitudes of $T_{\mathrm{C}}$ of both subsystem I and II.
There are two kinks seen on the mTRM, around 160 and 235~K.
These numbers tie nicely with the indications from the derivative $dM_{\mathrm{FM}}/dT$.
However, we can go a step further.
The recorded shape of mTRM can be quite accurately reproduced if we assume that the whole signal is composed of two Langevin functions $L(T)$, that is the $B_J(T)$ function obtained for $J = \infty$.
This is the classical limit for the molecular field theory, which is the best approximation to describe the $T$-dependence of magnetic moments of the precipitates \cite{Navarro-QA:2020_Materials,Gas:2021_JMMM_NWs,Navarro-QA:2021_SR}.
These two $L(T)$ and their sum are marked in Fig.~\ref{Fig-MT}~(a) by two gray dotted and a black dashed lines, respectively.
We find the agreement of the dashed line and the experimental envelope of mTRM as a very supportive to our interpretation, despite the fact that the magnitudes of $T_{\mathrm{C}}^{\mathrm{I}} \simeq  220$~K and $T_{\mathrm{C}}^{\mathrm{II}} \simeq  150$~K, used to generate the dotted-line components comprising the final dashed line, differ slightly from the magnitudes obtained from  the derivative of $dM_{\mathrm{FM}}(T)/dT$.


\begin{table}
	\caption{Selected magnetic compounds in Cr-Te system with ferromagnetic (T$_C$)
or antiferromagnetic (T$_N$) transition temperature and magnetic moment per Chromium ion.}		

\begin{tabular}{lrccl}\hline\hline \\*[-1ex]
Compound&T$_C$ (K)& T$_N$ (K) & $\mu_B$/Cr ion & Reference \\*[0.5ex] \hline\hline
CrTe &328& &   & \cite{Akram:1983_JMS} \\
Cr$_{\rm 0.6}$Te&237& & 3.51 &\cite{Liu:2017_PRB} \\
Cr$_{\rm 2}$Te$_{\rm 3}$&160, 180&130 & calc.: 3.03, 3.96; exp.: 2--2.7 & \cite{Chattopadhyay:1994_JournalPhaseEq},  \cite{Dijkstra:1989_JPCM}\\
              &198, 335&   &    & \cite{Konno:1993_JJAP}\\
Cr$_{\rm 3}$Te$_{\rm 4}$&315--340& & calc.: 3.32; exp.: 2.35 &  \cite{Chattopadhyay:1994_JournalPhaseEq}, \cite{Dijkstra:1989_JPCM}\\
Cr$_{\rm 5}$Te$_{\rm 8}$&180--230, 250& & 2.03 &  \cite{Luo:2018_JMMM}, \cite{Liu:2019_PRB}, \cite{Mondal:2019_JMMM}\\
 \hline\hline
\end{tabular}

\label{Table-CrTe}
\end{table}

The Table~\ref{Table-CrTe} lists chemical and magnetic characteristics of the most commonly met Cr$_m$Te$_n$ compounds. The lower $T_{\mathrm{C}}^{\mathrm{II}} \simeq  (160 \pm 10)$~K ties best with Cr$_2$Te$_3$.
This compound forms a  complex hexagonal unit cells with lattice parameters differing from that of PbTe: $a=6.814$~\AA, $c=12.073$~\AA  \cite{Andersen:1970_ActaChemScand}.
The layered structure of the unit cell accommodates Cr at three distinct positions and the strength of the Cr-Cr magnetic interaction depends sizably on their actual location \cite{Konno:1993_JJAP,Burn:2019_SR}.
Either a noncolinear arrangement of Cr moments \cite{Dijkstra:1989_JPCM} or even a partial moment compensation amongst these ions \cite{Youn:2007_JAP} are expected.
In our view the higher $T_{\mathrm{C}}^{\mathrm{I}} \simeq  (225 \pm 10)$~K ties favorably with Cr$_5$Te$_8$.
This is yet another example of a binary chromium telluride which assumes the NiAs-type structure.
However, here there are  four different crystallographic sites for Cr atoms \cite{Huang:2005_JSSC}  and also a noncolinear arrangement is likely, although FM behavior prevails \cite{Zhang:2018_JALCOM}.
This lack of the perfect Cr spins alignments in both Cr$_2$Te$_3$ and Cr$_5$Te$_8$ may give a good account of the steady increase of $M_{\mathrm{FM}}(H)$ as documented in Fig.~\ref{Fig-FM} at low and intermediate temperatures at high magnetic fields.

However, it has to bear in mind that the values presented in the Table were established for bulk compounds, that is the "free standing" forms of these compounds.
In the case of composite materials, like the sample studied here, the foreign nanocrystals (NCs) formed in  PbTe lattice differ in practically all material characteristics from the host, including the type and symmetry of their crystal structure and in lattice parameters.
Some NCs may form heretostructures coherent with the rock-salt structure, while the other do not.
Particularly, the latter are subjected to sizable strains.
These strains in turn are likely to modify their "native" magnetic characteristics including the spin texture and types of the spin-spin coupling.
Such effects have strong impact on the experimentally accessible properties, like the internal $T_{\mathrm{C}}$ of these NCs.
The possible  scale of this effect was demonstrated recently by documenting above 30\% increase of $T_{\mathrm{C}}$, from 315 to above 400~K, in MnAs NCs precipitated and strained in a nanotube geometry of wurtzite GaAs:Mn \cite{Kaleta:2019_NL}.
It has to be remarked, therefore, that the definitive identification of the chemical form of the NCs can only be possible by means of dedicated nanoscopic methods, like high resolution transmission electron microscopy, what is beyond the scope of this report.
Nevertheless, the results gathered so far and the discussion given above can provide an important guidance for further studies.


\begin{figure}[t!]
\begin{center}
\includegraphics[width = 8.0cm]{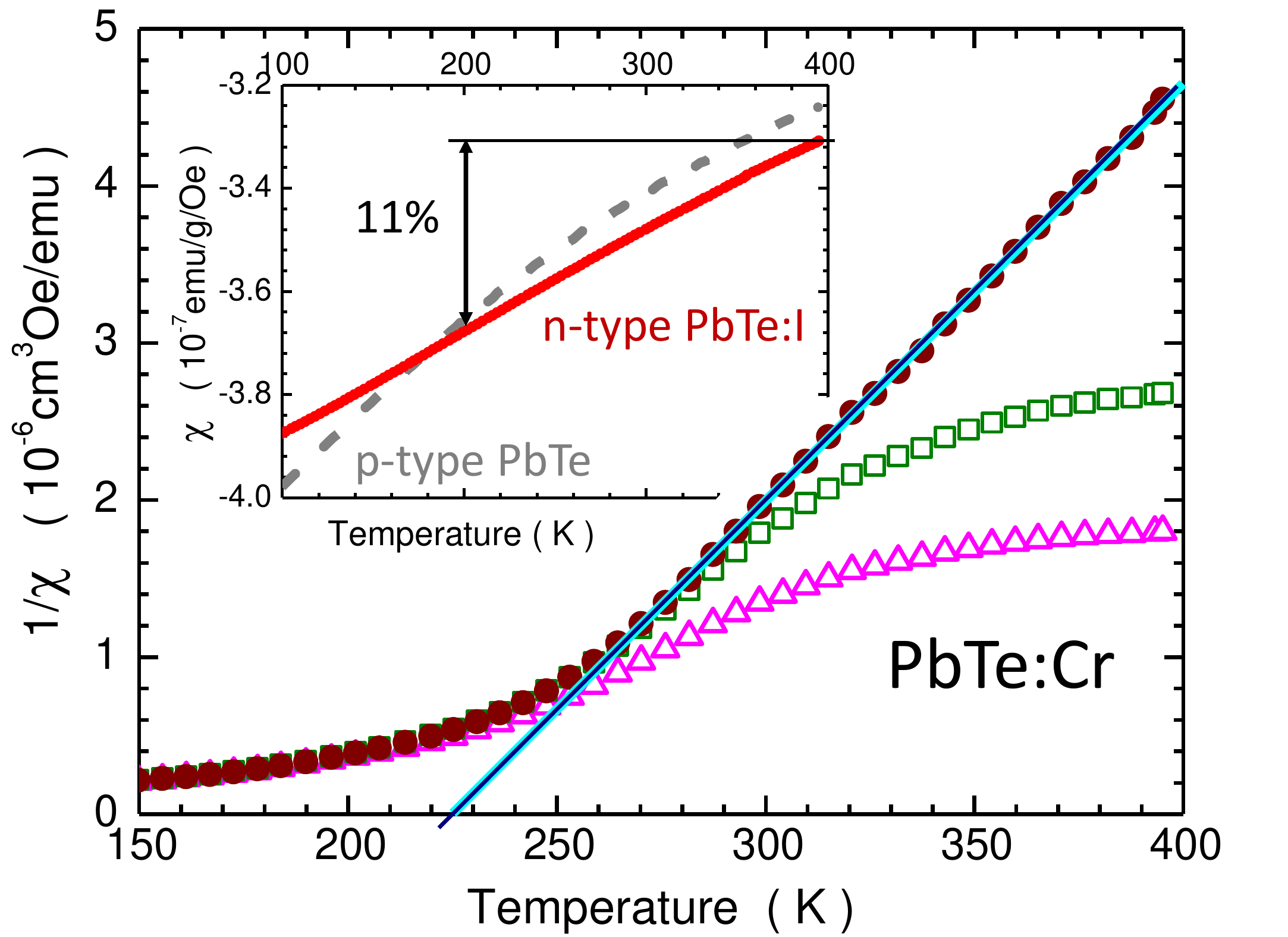}
\caption{(Color online)High temperature, $T$, dependence of the inverse of the magnetic susceptibility, $1/\chi$, of the PbTe:Cr sample studied here (symbols). Triangles exemplify results obtained when a $T$-independent $\chi_{\mathrm{PbTe}}$ is assumed. Squares mark results after removal of the paramagnetic contribution from Cr$^{3+}$ ions. The bullets denote the final magnitudes of $1/\chi$ after accounting for the $T$-dependence of $\chi$ in an n-type PbTe:I sample of a comparable concentration of electrons: $n \simeq 10^{19}$~cm$^{-3}$. The solid line marks the Curie-Weiss law. Inset: a comparison of the $T$-dependence of $\chi$ in p-type (gray dashed line) and n-type (red solid line) PbTe. The double pointing arrow represents the range of changes of $\chi$ in the relevant here $T$-range $200 < T < 400$~K.  }
\label{Fig-CW}
\end{center}
\end{figure}
Finally, we establish the number of the ferromagnetically coupled Cr atoms, $x_{\mathrm{Cr}}^{\mathrm{FM}}$, that is these atoms which precipitated into the FM-like precipitates.
This assessment is accomplished  by the analysis of the high temperature part of the magnetic susceptibility, $\chi$, within the frame of the Curie-Weiss approach.
The main obstacle in this, otherwise straightforward, procedure is the necessity of proper identification, evaluation, and removal of other contributions to the high-$T$ magnetic signal of the sample.

If we assume, as it is ubiquitously done \cite{Hedgcock:1985_CanJP,Gorska:1988_PRB,Skipetrov:2002_PRB,Bain:2008_JCE,Nielsen:2012_PRB,Podgorni:2015_JALCOM}, a $T$-independent diamagnetism of host PbTe, than a rather rounded shape of $\chi^{-1}(T)$ is obtained - open triangles in Fig.~\ref{Fig-CW}.
It is definitely too rounded for a reasonable analysis.
We take here $\chi_{\mathrm{PbTe}} \simeq -3.4 \times 10^{-7}$~emu/g/Oe after ref.~\cite{Hedgcock:1985_CanJP}, however the same convex curvature of $\chi^{-1}(T)$ remains for other magnitudes of $\chi_{\mathrm{PbTe}}$, what informs us about the existence of other unaccounted contributions.

We identify the first undesirable contribution to $\chi$ to be a high-$T$ tail of the PM signal of the  Cr$^{3+}$  and Cr$^{2+}$  ions.
The PM component has been found to contribute to about 2/3 of the total low-$T$ magnetization in this sample, so its contribution to $\chi$ at high $T$ cannot be a negligible one.
We calculate and subtract the relevant high-$T$ magnitude of $\chi_{\mathrm{PM}}(T)$ using the concentrations of $x_{\mathrm{Cr^{3+}}}$ and $x_{\mathrm{Cr^{2+}}}$ evaluated previously.
The PM-ions-corrected $\chi^{-1}(T)$ is marked by open squares in Fig.~\ref{Fig-CW}.
An improvement is obvious, but a good deal of the disturbing concave rounding remains.

At this point we recall the fact that PbTe is a narrow-gap semiconductor and its susceptibility is rather substantially $T$-dependent.
As a matter of fact, the absolute values of $\chi_{\mathrm{PbTe}}$ depend also on the type of dopants and on the  doping level.
To establish the relevant here $T$-dependence of $\chi_{\mathrm{PbTe}}$ we perform direct magnetic measurements of two PbTe samples: a nominally undoped one, that is a p-type material, and an n-type PbTe:I. The doping level in the latter of about $10^{19}$~cm$^{-3}$ corresponds to the n-type doping level brought about by $x_{\mathrm{PM}}$ in our sample.
Both compounds are grown using the same Bridgman method as the PbTe:Cr sample.
The results are given in the inset to Fig.~\ref{Fig-CW}. 
It is very instructive to note that both values are $T$-dependent and that these dependences are rather substantial.
In particular $\chi_{\mathrm{PbTe:I}}$ changes its magnitude by about 11\% in the relevant here $T$-range ($200 < T < 400$~K), so it proves absolutely essential to take into account the real $T$-dependence of the susceptibility of the host narrow gap semiconductor lattice.

The solid circles in Fig.~\ref{Fig-CW} represent the final magnitudes of $\chi^{-1}(T)$ corresponding to the Cr atoms conglomerated in Cr$_m$Te$_n$ precipitates in our sample, that is after correcting the raw $m(T)$ data for $T$-dependent: (i) paramagnetism of the substitutional Cr ions and (ii) the experimental magnitudes of the $T$-dependent diamagnetism of n-type PbTe.
The resulting dependency can be unambiguously approximated by a straight line.
Its slope, $C \simeq 26500$,  following the Curie-Weiss law, yields the sought concentration of Cr  in the Cr-rich NCs, $x_{\mathrm{Cr}}^{\mathrm{NC}}= 3 k_{\mathrm{B}}[C N_0 g^2 \mu^2_{\mathrm{B}} S(S+1)]^{-1}$, or its approximate magnitude, since the value of $S$ (or of the  effective magnetic moment) of Cr atoms in the  formed here Cr$_m$Te$_n$ compounds is known at present only approximately.
Being guided by the numbers collected in Table~1 we obtain $x_{\mathrm{Cr}}^{\mathrm{NC}} \simeq 0.08$\% by taking $S=3/2$, or $x_{\mathrm{Cr}}^{\mathrm{NC}} \simeq 0.05$\% for $S=2$.
Particularly the latter estimate ties remarkably well with our former estimation of $x_{\mathrm{Cr}}^{\mathrm{NC}} \simeq 0.04$\% from the comparison of the magnitudes of the PM and FM-like components, presented in  Fig.~\ref{Fig-FM}.
Remarkably, the latter result is obtained from the analysis of the $M(H,T)$ data obtained at complementary conditions to those of Fig.~\ref{Fig-CW}, namely at low and moderate temperatures and at high magnetic fields.


We close our discussion by mentioning that the quantitative analysis of the magnetic data in the wide range of magnetic field  and temperature yields separately the total concentrations of the paramagnetic (substitutional) Cr and  ions,  $x_{\mathrm{PM}}/N_0 \simeq 0.08$\%  and of those atoms which precipitated in Cr-rich nanocrystals, $x_{\mathrm{Cr}}^{\mathrm{NC}}/N_0 = 0.05$ to 0.08\%, totaling to 0.13 -- 0.16\%  per cation site.
We indicate these values of $x$ in Fig.~\ref{Fig-sklady} finding that the magnetic determination yields the total magnitude of Cr content to be close to but markedly below the concentration determined for this part of the ingot by the energy dispersive x-ray fluorescence analysis.

\section{Conclusion}

Detailed temperature and magnetic field dependence of magnetization of IV-VI semiconductor PbTe doped with mixed valence transition metal Cr$^{2+/3+}$ has been studied by SQUID magnetometry in order to quantitatively determine the contribution of substitutional  Cr$^{3+}$ or  Cr$^{2+}$ ions as well as various Cr-Te magnetic inclusions.
The analysis yields the total concentration of Cr$^{2+/3+} \simeq 1.1 \times 10^{19}$~cm$^{-3}$, that is it confirms the existence of paramagnetic  Cr ions acting as resonant donors, which are known to improve the thermoelectric figure of merit parameter zT of this  semiconductor.
This result provides a new experimental support for the proposed model of the electronic structure of PbTe:Cr system with resonant  Cr$^{2+/3+}$ state located (at low temperatures) about 100 meV above the bottom of the conduction band.
The analysis of temperature dependence of the magnetization resolved also  two ferromagnetic-like contributions that we assign to Cr$_2$Te$_3$ and  Cr$_5$Te$_8$. Both compounds form precipitates in the PbTe host lattice and,  as an ensemble, exhibit (blocked) superparamagnetic properties.
The temperature dependence of the magnetic susceptibility of both n- and p-type PbTe in the temperature range between 150 and 400~K proved essential in the evaluation of the correct magnitudes of the total number of Cr ions present in the Cr-rich precipitates.
The total concentration of Cr atoms established in the material on the account of the magnetic studies is  found to compare favorably to that obtained upon the energy dispersive x-ray fluorescence analysis.
The whole body of the evidences presented in the report establishes fine magnetometric studies as a complementary and remarkably accurate characterization tool of composite materials, proving itself to be a first-hand selection tool of prospective materials for the studies of thermoelectricity in magnetic materials.

\section*{Acknowledgments}
This study has been supported by the National Science Centre (Poland) through project OPUS (UMO - 2017/27/B/ST3/02470) and by the National Science Centre for Development (Poland) through grant TERMOD No TECHMATSTRATEG2/408569/5/NCBR/2019 and by the Foundation of Polish Science through the IRA Programme co-financed by EU within SG OP.
The authors acknowledge fruitful discussions with Hanka Przybyli\'nska.



\end{document}